\documentclass[seceq]{jpsj2} 
\usepackage{color}
\title{Infectious Default Model with Recovery and Continuous Limits}

\author{Ayaka \textsc{Sakata}$^{1}$\thanks{E-mail address:
sakata@complex.c.u-tokyo.ac.jp}, 
Masato \textsc{Hisakado}$^{2}$\thanks{E-mail address:
hisakado\_masato@standardandpoors.com} and
Shintaro \textsc{Mori}$^{3}$\thanks{E-mail address: mori@sci.kitasato-u.ac.jp}}

\inst{$^{1}$Department of Pure and Applied Science,
University of
Tokyo,\\
 Komaba, Meguro-ku, Tokyo 153-8902\\
$^{2}$Standard \& Poor's,
Marunouchi, Chiyoda-ku, Tokyo
100-0005 \\
$^{3}$Department of Physics, School of Science,
Kitasato University,\\
 Kitasato, Sagamihara,
Kanagawa 228-8555 
}

\abst{We introduce an infectious default and recovery model for $N$
obligors. The obligors are assumed
to be exchangeable and their states are described by $N$
Bernoulli-type random variables $S_{i} (i=1,\cdots,N)$.
They are expressed by multiplying independent Bernoulli variables
$X_{i},Y_{ij}$ and $Y^{\prime}_{ij}$, and the default and recovery infections
are described by $Y_{ij}$ and $Y^{\prime}_{ij}$.
We obtain the default probability function $P(k)$ for $k$ defaults. 
By considering a continuous limit,
we find two nontrivial probability distributions with a reflection
symmetry of $S_{i} \leftrightarrow 1-S_{i}$. Their profiles are singular 
and oscillating and 
we theoretically investigate it.
We also compare $P(k)$
with an implied default distribution
function inferred from the quotes of iTraxx-CJ,
which is a portfolio credit derivative of Japanese 50 companies.
In order to explain the behavior of  the implied distribution,
the recovery effect may be necessary. 
}

\kword{default correlation, correlated binomial, default distribution,
continuous limit}

\begin{document}
\maketitle

\section{Introduction}
The cooperative phenomena,
especially phase transitions, have been extensively studied and 
continue to be important subjects until today. They have provided
universal paradigm for physics, sociology, and economy. 
The economical systems composed of a large number of interacting units
have been studied from this viewpoint. \cite{MS,CMZ}
Recently, systemic failure problems are being focused upon in econophysics,
\cite{AH,IJ,LH,KSY,MTT}
financial engineering, \cite{DL,KMH,HKM,MKH2,MKH} and computer
engineering, \cite{Ba}
  and many probabilistic models have been proposed.
This has been motivated by the fact that the description of systemic failures is necessary
to control and manage them. Another motivation is that credit risk 
markets are now growing, and the pricing of products 
is an immediate concern.\cite{Sch} For this purpose, it is necessary to
develop probabilistic models that can describe credit risks.
 
The difficulty in the description of systemic failures arises from the fact that 
they are not independent events. If they are independent, the description
is very easy and we only need Bernoulli-type random variables $S_{i}$ 
denoting the element $i$'s failure or not by $S_{i}=1$ or $S_{i}=0$ respectively. 
However, there are many 
phenomena wherein the ``correlation'' between the failure events is very
important. For example, in a network of storage systems, if
a node fails, the failure can propagate to other nodes. 

In credit risk markets, the same type of risk propagation
is found to occur. A percolation-type probabilistic model was proposed
to describe bank bankruptcies, where interbank deposits lead to
collective credit risks. The probability of $l$ failures obeyed 
the power law $P(l)\sim l^{1-\tau}$ with the Fisher exponent $\tau$ at its
critical point. In a study,\cite{DL} a default infection
mechanism was proposed to describe the risk-dependency structures.
The constituents are obligors, and the risk is whether he (or she)
can refund before the expiry date. Such a risk is called a default risk.
Davis and Lo introduced independent Bernoulli-type random 
variables $Y_{ij}$, which describes the infection from a bad obligor
$j$ to a good one $i$. They explicitly obtained the probability function for $k$ defaults,
$P(k)$. They estimated the effect of the default correlation
on $P(k)$.  

One of the crucial problems with these studies 
is that their it is difficult to describe whether $P(k)$ values
do describe the empirical default distribution $P(k)$ or not. 
Because of the relative scarcity of good data on credit events,
it was impossible to compare the models by using the empirical 
data. Recently, from the market quotes on credit risk products, 
it becomes possible to infer the default distribution function.\cite{MKH,HW} 
We can compare and calibrate the probabilistic models.
In the present paper, we generalize the model proposed by Davis and Lo
by introducing a recovery effect. We compare the default distribution
$P(k)$ with an implied value of that of the credit market and calibrate the model 
parameters. With regard to the bulk shape, we see that the calibrated $P(k)$ value looks 
similar.

The outline of the present paper is as follows.
Section \ref{model} gives a brief introduction to the infectious default 
model proposed by Davis and Lo,\cite{DL}
and we modify it by introducing a recovery process.
We obtain the default probability function $P(k)$ for $k$ defaults.
In section \ref{Cont}, we take the continuous limit of $P(k)$ 
with finite $P_d$ and non zero correlation $\rho>0$.
We find two non trivial probability distribution functions
with a reflection symmetry.
They exhibit oscillating behaviors and we investigate the mechanism. 
We compare the model distribution function $P(k)$
with the market implied function in section \ref{Comp}. 
Section \ref{Conc} summarizes our results and future problems are discussed.

\section{Infectious default model}\label{model}
We consider $N$ exchangeable obligors whose states are described by
random variables $S_{i}(i=1,2,\cdots ,N)$ such
that $S_{i}=1$ if obligor $i$ defaults and $S_{i}=0$ otherwise. 
Here, the term ``exchangeable'' means 
the non-dependency of the joint probabilities 
$P(S_{1},S_{2},\cdots,S_{N})$ on the exchange of 
$S_{i}\leftrightarrow S_{j}$ for any pair of $(i,j)$.
The number of defaults is 
\begin{equation}
K=S_{1}+S_{2}+\cdots+S_{N}.
\end{equation}
The value of $S_{i}$ is determined as follows.
For $i=1,\cdots,N$ and $j=1,\cdots,N$ with $j\neq i$, let $X_{i},Y_{ij}$ be
independent Bernoulli-type random variables with probability function
\begin{eqnarray}
\nonumber
\mbox{Prob.}[X_{i}=1]&=&p ,\\
\mbox{Prob.}[Y_{ij}=1]&=&q .
\label{pq}
\end{eqnarray}
$S_{i}$ are defined as
\begin{equation}
S_{i}=X_{i}+(1-X_{i})(1-\Pi_{j\neq i}(1-Y_{ij}X_{j})).
\label{Si}
\end{equation}
Here, $X_{i}$ is the internal state 
variable which describes whether the obligor is in a good state ($X_{i}=0$)
or not ($X_{i}=1$). 
$S_{i}$ is also the state variable that describes
whether the obligor is defaulted ($S_{i}=1$) or not ($S_{i}=0$), 
this is determined by not only the internal state
but also the external environment.
If $X_{i}=1$, $S_{i}=1$ and the obligor is defaulted.
Even if $X_{i}=0$, 
obligor $i$ can be defaulted.
$Y_{ij}$ represents the influence of another bad obligor
($X_{j}=1$) on obligor $i$. 
A default infection from a bad obligor $j$
takes place if $Y_{ij}=1$ and $X_{j}=1$. 
$S_{i}$ becomes 1 and the obligor is defaulted.
The second term of eq.(\ref{Si}) represents this effect.

We introduce a supporting effect from other good obligors
in addition to
the default infection.
In fact, it may occur that 
a good  obligor supports other bad obligors and
the latter can circumvent their defaults.
We introduce new independent
Bernoulli-type random variables $Y^{\prime}_{ij}$ in addition to eq.(\ref{pq}).
For $i=1,\cdots,N$ and $j=1,\cdots,N$ with $j\neq i$, they 
have the probability function 
\begin{equation}
\mbox{Prob.}[Y_{ij}^{\prime}=1]=q^{\prime}.
\end{equation}
We introduce the following model equation for $S_{i}$: 
\begin{equation}
S_{i}=X_{i}\Pi_{j\neq
 i}(1-Y_{ij}^{\prime}(1-X_{j}))+(1-X_{i})(1-\Pi_{j\neq
 i}(1-Y_{ij}X_{j})).
\label{S_S}
\end{equation}
Eq.(\ref{S_S}) reveals that
even when $X_{i}=1$, if 
$X_{j}=0$ and $Y^{\prime}_{ij}=1$,
obligor $i$ is supported by obligor $j$ and avoids being defaulted.
We note that eq.(\ref{S_S}) has a default, non-default symmetry. 
We get $1-S_{i}$ by
substituting $X_{i}\to 1-X_{i}$ and $Y_{ij}\leftrightarrow  Y^{\prime}_{ij}$.
This model can be reduced to the original infectious default 
model by substituting 
$Y'_{ij}=0$ into eq.(\ref{S_S}).

The probability distribution function $P(k)$ for $k$ defaults 
is given by
\begin{equation}
P(k)=\mbox{Prob.}[K=k]=
{_{N}C_{k}}\times\sum_{l=0}^{k}
\sum_{m=0}^{N-k}\alpha_{N,k}^{p,q,q^{\prime}}(l,m)
\label{PK}
\end{equation}
where
\begin{eqnarray}
\nonumber
\alpha_{N,k}^{p,q,q^{\prime}}(l,m)&=&_{k}C_{l}\times{_{N-k}C_{m}}\times p^{N-k-m+l}(1-p)^{k-l+m}\\
\nonumber
&&\ \ \times (1-q^{\prime})^{l(k+m-l)}(1-q)^{m(N-k-m+l)}\\
&&\ \ \times
 (1-(1-q)^{N-k-m+l})^{k-l}(1-(1-q^{\prime})^{k+m-l})^{N-(k+m)}.
\label{alpha}
\end{eqnarray}
\begin{figure}[tb]
\begin{center}
\includegraphics[scale=0.4]{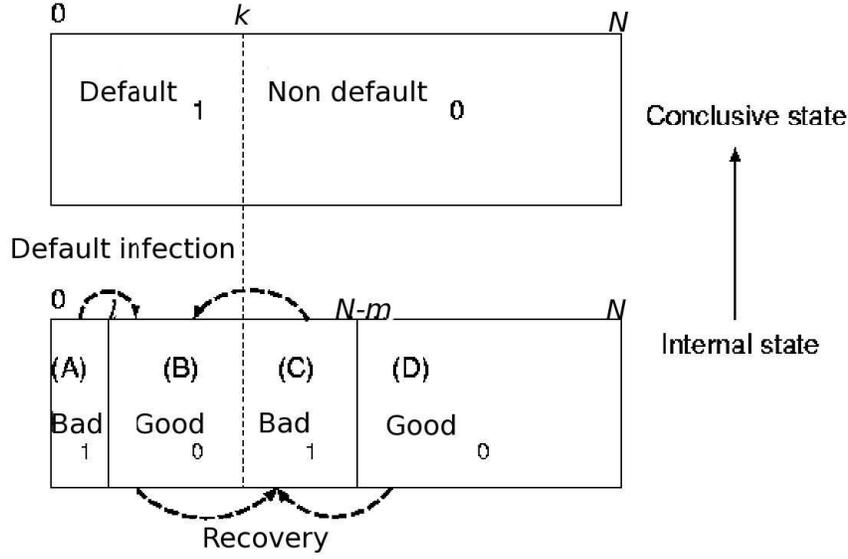} 
\caption{Pictorial representation from internal states $\{X_{i}\}$ to
a conclusive state of $k$ defaults and $N-k$ non-defaults.}
\label{lm}
\end{center}
\end{figure} 
We explain the derivation of eq.(\ref{alpha}). In Fig.1,
there are $N$ obligors.
$k$ obligors are defaulted and $N-k$ obligors are
non-defaulted.
The $k$ defaulted obligors are classified into
two categories: (A) and (B).
(A) contains $l$ bad obligors, which are never supported by other
good obligors.
(B) contains $k-l$ good obligors, which are
infected by other bad obligors and thus get
defaulted. The number of different possible combinations of $l$ items
from $k$ different items is $_{k}C_{l}$.
Further, there are two categories (C) and (D)
for the $N-k$ non-defaulted obligors.
(C) contains $N-k-m$ bad
obligors and (D) contains $m$ good obligors.
The $N-k-m$ bad obligors are supported by
other good obligors and they are prevented from being defaulted.
The $m$ good obligors are never infected to be defaulted. 
The number of different possible combinations of $m$ items
from $N-k$ different items is $_{N-k}C_{m}$.
In other words, the conclusive $k$ defaults and $N-k$ non-defaults
are made from the $N-k-m+l$ bad obligors
and $k-l+m$ good obligors in the internal configuration by the infection and recovery mechanism.
The internal configuration is realized with probability
$p^{N-k-m+l}(1-p)^{k-l+m}$.
$l$ bad obligors among the $N-k-m+l$ obligors are not supported by the 
$k-l+m$ good obligors; this probability is given by
$(1-q^{\prime})^{(k-l+m)l}$.
$m$ good obligors are never infected by $N-k-m+l$ bad obligors;
this probability is given by $(1-q)^{m(N-k-m+l)}$.
$k-l$ good obligors must be infected by $N-k-m+l$ bad obligors;
this probability is given by $(1-(1-q)^{N-k-m+l})^{k-l}$.
$N-k-m$ bad obligors must be supported by $k-l+m$ good obligors;
this probability is given by $(1-(1-q^{\prime})^{k+m-l})^{N-k-m}$.
Therefore, the probability of $k$ defaults and $N-k$ non-defaults
from a configuration $(l,m)$ is given by
$\alpha_{N,k}^{p,q,q^{\prime}}(l,m)$, as shown
in eq.(\ref{alpha}). We obtain
$P(k)$ as the summation of $\alpha_{N,k}^{p,q,q^{\prime}}(l,m)$ over $l,m$.

The expected value of the number of defaults $K$ is
\begin{equation}
<K>=N[p(1-q^{\prime}(1-p))^{N-1}+(1-p)(1-(1-qp)^{N-1})],
\end{equation}
and the default probability $P_{d}$ is given as
\begin{equation}
P_d=<K>\slash N=p(1-q^{\prime}(1-p))^{N-1}+(1-p)(1-(1-qp)^{N-1}).
\label{P_S}
\end{equation}
The variance is
\begin{equation}
\sigma ^{2}_{K}=<K>+N(N-1)\beta_{N}^{p,q,q^{\prime}}-<K>^{2},
\end{equation}
where
\begin{eqnarray}
\nonumber
\beta_{N}^{p,q,q^{\prime}}&=&<S_{i}S_{j}>\\
\nonumber
&=&p^{2}\{1-2q^{\prime}(1-p)+q'^{2}(1-p)^{2}\}^{N-2}\\
\nonumber
&&\ \ +2p(1-p)(1-q^{\prime})\{(1-q^{\prime}(1-p))^{N-2}\\
\nonumber
&&\hspace{1cm} -(1-q)(1-q^{\prime}(1-p)-pq)^{N-2}\}\\
&&\ \ +(1-p)^{2}[1-2(1-pq)^{N-2}+(1-2pq+pq^{2})^{N-2}]
\label{beta}
\end{eqnarray}
and the correlation coefficient is given by
\begin{equation}
\rho=\frac{\beta_{N}^{p,q,q^{\prime}}-P_{d}^{2}}{P_d(1-P_d)}.
\label{cor_S}
\end{equation}
 
\begin{figure}[tb]
\hfill
\begin{minipage}[H]{.45\textwidth}
\begin{center}
\includegraphics[scale=0.2]{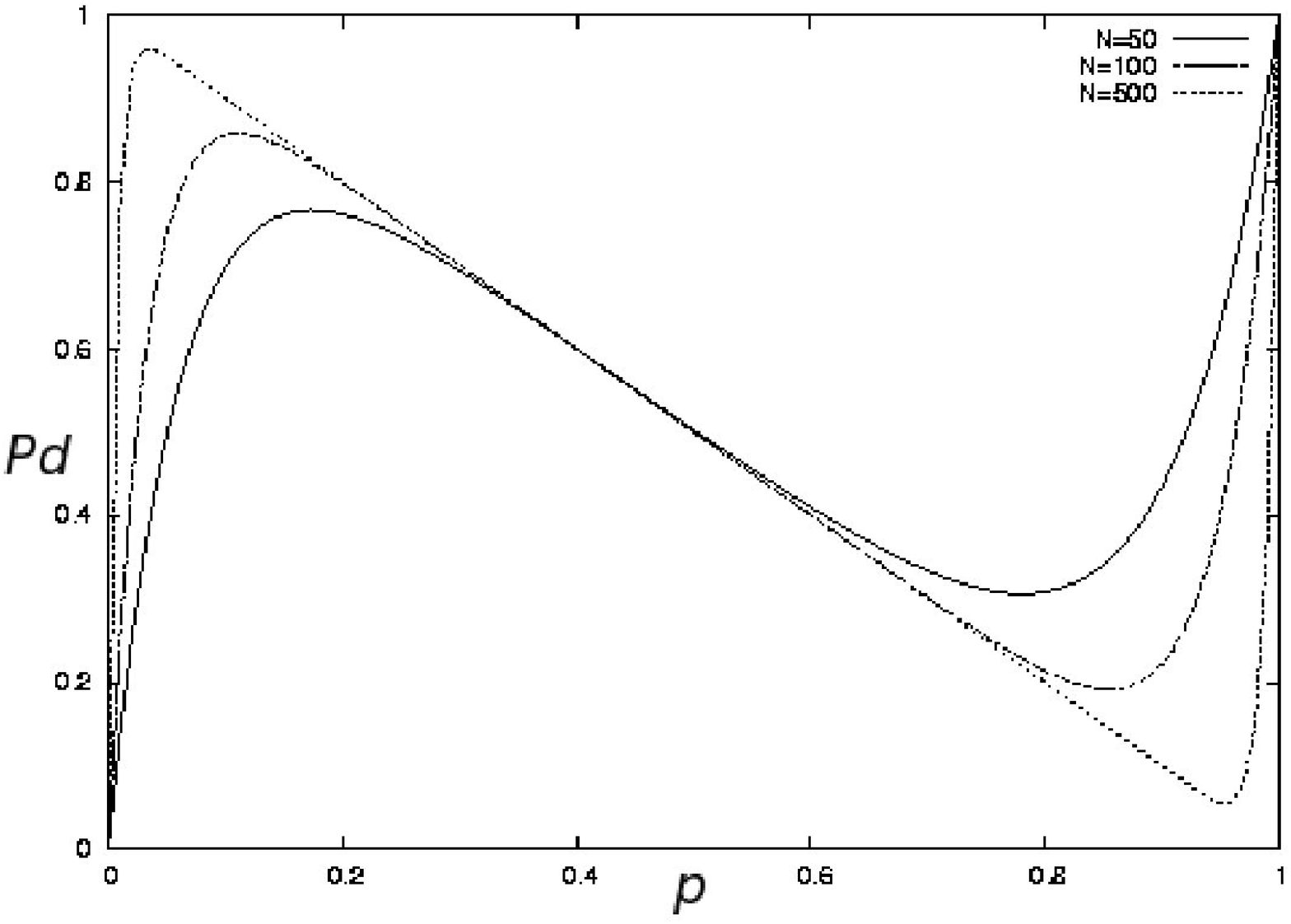}
\caption{Plot of $P_{d}$ vs. $p$. We set $N=50,100$, and $500$,
 $q=0.3$, and $q^{\prime}=0.2$.\\ }
\label{cubic}
\end{center}
\end{minipage}
\hfill
\begin{minipage}[H]{.45\textwidth}
\begin{center}
\includegraphics[scale=0.2]{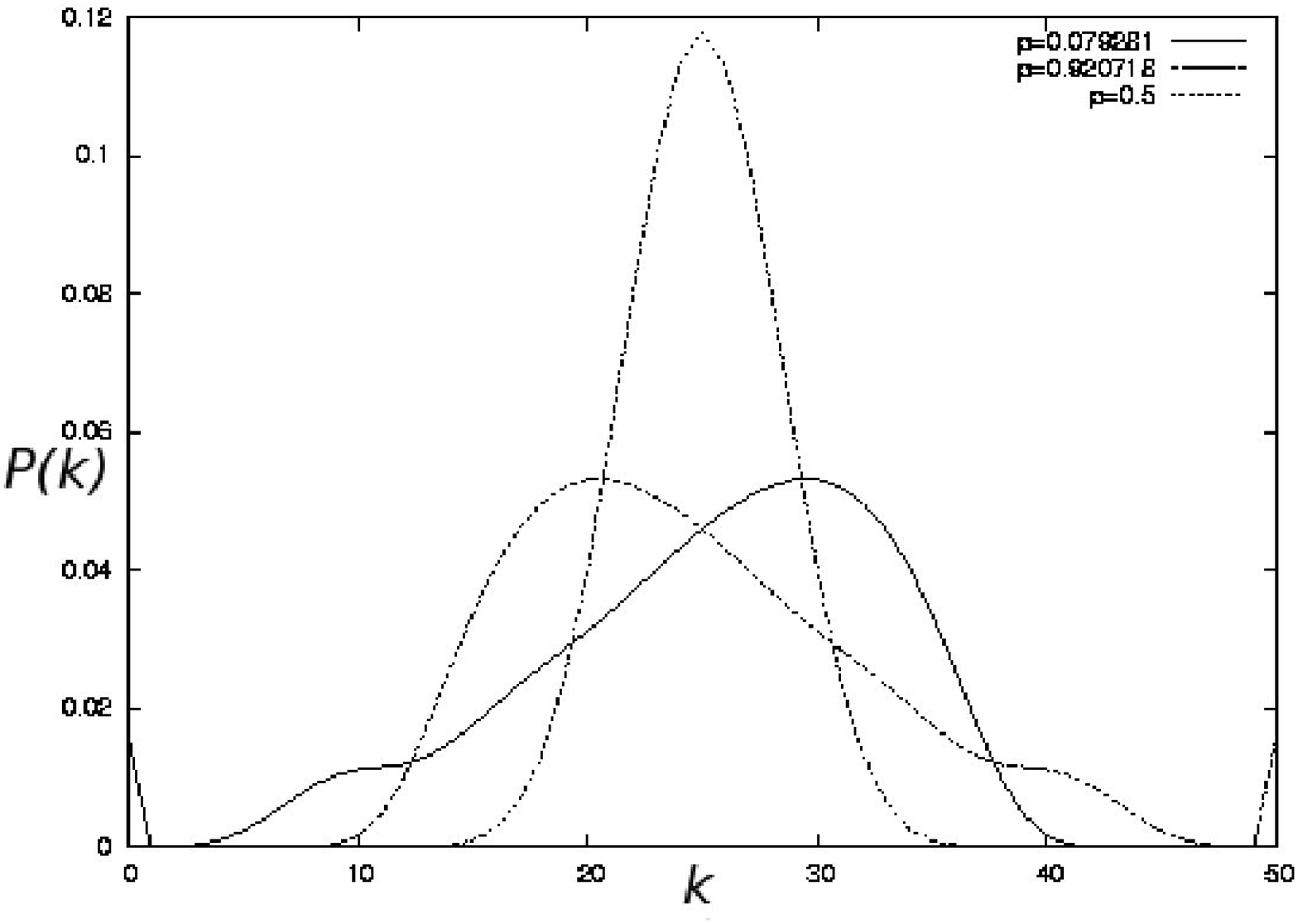} 
\caption{Plot of $P(k)$ for $P_{d}=0.5$ and $q=q^{\prime}=0.2$.
The left, middle, and right solutions are plotted.}
\label{P_S_N50}
\end{center}
\end{minipage}
\hfill
\end{figure}

We find that there are multiple solutions $(p,q,q^{\prime})$ corresponding to
a value of $P_{d}$.
In particular, for large $N$, there are three solutions.
For example, there are three solutions 
$p=0.808310, 0.5$, and $0.191680$
for 
$N=100,P_{d}=0.5$, and $q=q^{\prime}=0.05$. On the other hand,
there is only one solution $p=0.5$
for $N=50,P_{d}=0.5$, and $q=q^{\prime}=0.05$.
This is because 
for arbitrary $q,q^{\prime}\ne 0$, $P_{d}$ 
behaves as that in Fig.\ref{cubic}.
$P_{d}(p,q,q^{\prime})$ starts from 0 at $p=0$
to 1 at $p=1$.
For intermediate values of $p$,
$P_{d}$ rapidly increases to 1 and then decreases to 0 near $p=1$ in the
large $N$ limit.
Thereafter, $P_{d}$ again increase rapidly to 1 toward $p=1$.
Such a behavior can be explained by eq.(\ref{P_S}).
There are three $p$ solutions corresponding to a $P_{d}$ value;
they are referred to as 
left, middle and right solutions
according to the order of $p$.
The parameter region $(q,q')$ in which there are three solutions of $p$ 
expands with $N$; further, for the limit $N\to\infty$, 
it covers the entire parameter space $(q>0,q'>0)$.

The profiles of the three solutions are shown in Fig.\ref{P_S_N50}.
We set $N=50, P_{d}=0.5$, and $q=q^{\prime}=0.2$. 
The three solutions are realized at $p=0.079281$ (left), 
$p=0.5$ (middle), and $p=0.920719$ (right).
The profiles of the probability distribution functions of the left 
and right solutions are reflection symmetric.
The origin of the symmetry arises from the reflection symmetry of 
eq.(\ref{S_S}). 
$P(k)$ for the middle solution ($p=0.5$) has a 
symmetrical profile and  is almost a
binomial distribution $\mbox{Bi}(50,0.5)$.

\section{Continuous limit and probability distribution function}
\label{Cont}
In this section,
we consider the continuous limit of eq.(\ref{alpha}).
It is required to take the limit $N\to\infty$ 
with non zero correlation because the probability distribution
function of the uncorrelated variables is a binomial distribution.
Its continuous limit reduces to a trivial delta function.
We need to consider the continuous limit with fixed $P_{d}$ and $\rho$.
Writing explicitly, 
$\rho=(<S_{i}S_{j}>-<S_{i}><S_{j}>)\slash P_{d}(1-P_{d})$ is calculated as 
\begin{eqnarray}
\nonumber
\rho&=&\{p^{2}[(1+(q'^{2}-2q^{\prime})(1-p))^{N-2}-(1-2q^{\prime}(1-p)+q'^{2}(1-p)^{2})^{N-1}]\\
\nonumber
&&\ \ +(1-p)^{2}[(1+(q^{2}-2q)p)^{N-2}-(1-2qp+q^{2}p^{2})^{N-1}]\\
\nonumber
&&\ \
-2p(1-p)[(1-p)q(1-pq)^{N-2}-pq^{\prime}(1-q^{\prime}(1-p))^{N-2}\\
\nonumber
&&\ \ \ \ \ +(1-q^{\prime})(1-q)(1-qp-q^{\prime}(1-p))^{N-2}\\
&&\ \ \ \ \ -(1-q^{\prime}(1-p))^{N-1}(1-qp)^{N-1}]\}\slash P_{d}(1-P_{d}),
\label{rho_}
\end{eqnarray}
where
\begin{equation}
P_{d}=p(1-q^{\prime}(1-p))^{N-1}+(1-p)(1-(1-qp)^{N-1}).
\label{Pd}
\end{equation}
There are three terms in eq.(\ref{rho_}).
The first term is derived from $<X_{i}X_{j}>$, which is proportional to
$p^{2}$. The second term is derived from $<(1-X_{i})(1-X_{j})>$, which is
proportional to $(1-p)^{2}$.
The last term is derived from $<X_{i}(1-X_{j})>$ and $<(1-X_{i})X_{j}>$,
which is proportional to $2p(1-p)$.
At least one term must be non-zero in the continuous limit
in order to retain the correlation.
In order to fix $P_{d}$ in the limit
$N\to\infty$, 
it is necessary
to set the parameters $p\times q$ or $(1-p)\times q^{\prime}$
to be proportional to $1\slash N$
due to the presence of the $N$th power in  eq.(\ref{Pd}).
To satisfy this condition, we must set $p, q, q^{\prime}$ 
such that the non zero correlation is maintained. 
Using a proportional coefficient $\alpha$,
if we set $p=\alpha\slash N$,
the resulting expression corresponds to the left solution in the previous section.
The correlation is maintained due to
the first term of eq.(\ref{rho_}).
For $p=1-\alpha\slash N$, 
the resulting expression corresponds to the right solution;
the second term of eq.(\ref{rho_}) remains.
Instead, if we set $q,q^{\prime}\propto 1\slash N$ and $p$ to be finite, 
eq.(\ref{rho_}) vanishes
in the limit $N\to\infty$ and the correlation disappears.
The resulting expression corresponds to the middle solution in the previous section
and the model becomes a binomial distribution.

In the above limit, $P_{d}$ and $\rho$ can be easily estimated. 
We set $p=\alpha\slash N$ and substitute this value in eq.(\ref{rho_})
and eq.(\ref{Pd}). We obtain
\begin{equation}
P_{d}=1-e^{-\alpha q},
\label{P_left}
\end{equation}
\begin{equation}
\rho=\frac{e^{-\alpha q}(e^{\alpha q^{2}}-1)}{1-e^{-\alpha q}}.
\label{rho_left}
\end{equation}
If we set $p=1-\alpha\slash N$, we get
\begin{equation}
P_{d}=e^{-\alpha q^{\prime}}
\label{P_right}
\end{equation}
and
\begin{equation}
\rho=\frac{e^{-\alpha q^{\prime}}(e^{\alpha q'^{2}}-1)}{1-e^{-\alpha q^{\prime}}}.
\label{rho_right}
\end{equation}
We see that
the above two non trivial solutions can retain their non-zero
correlations
in the continuous limit.
From the symmetric property of the model, we note that
eq.(\ref{P_right}) and eq.(\ref{rho_right}) can be derived by
the substituting
$P_{d}\leftrightarrow 1-P_{d},q\leftrightarrow q^{\prime}$, and
$p\leftrightarrow 1-p$ in eq.(\ref{P_left}) and eq.(\ref{rho_left}). 

\begin{figure}
\hfill
\begin{minipage}[H]{.45\textwidth}
\begin{center}
\includegraphics[scale=0.2]{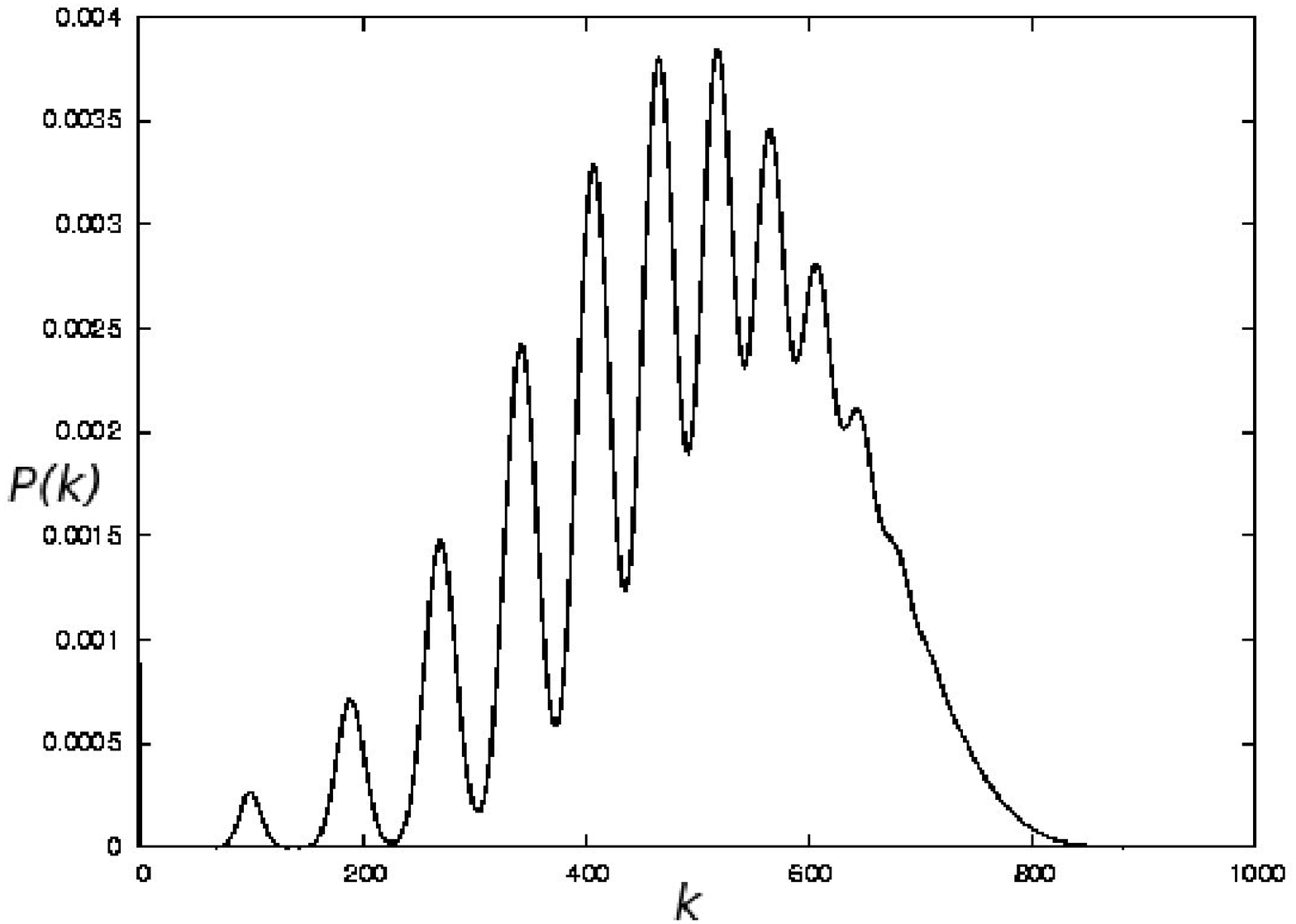} 
\caption{Plot of $P(k)$ for $P_{d}=0.5,q=q^{\prime}=0.1,N=1000,\\p=\alpha\slash N=0.007$, and $\rho=0.071773$}
\label{left}
\end{center}
\end{minipage}
\hfill
\begin{minipage}[H]{.45\textwidth}
\begin{center}
\includegraphics[scale=0.2]{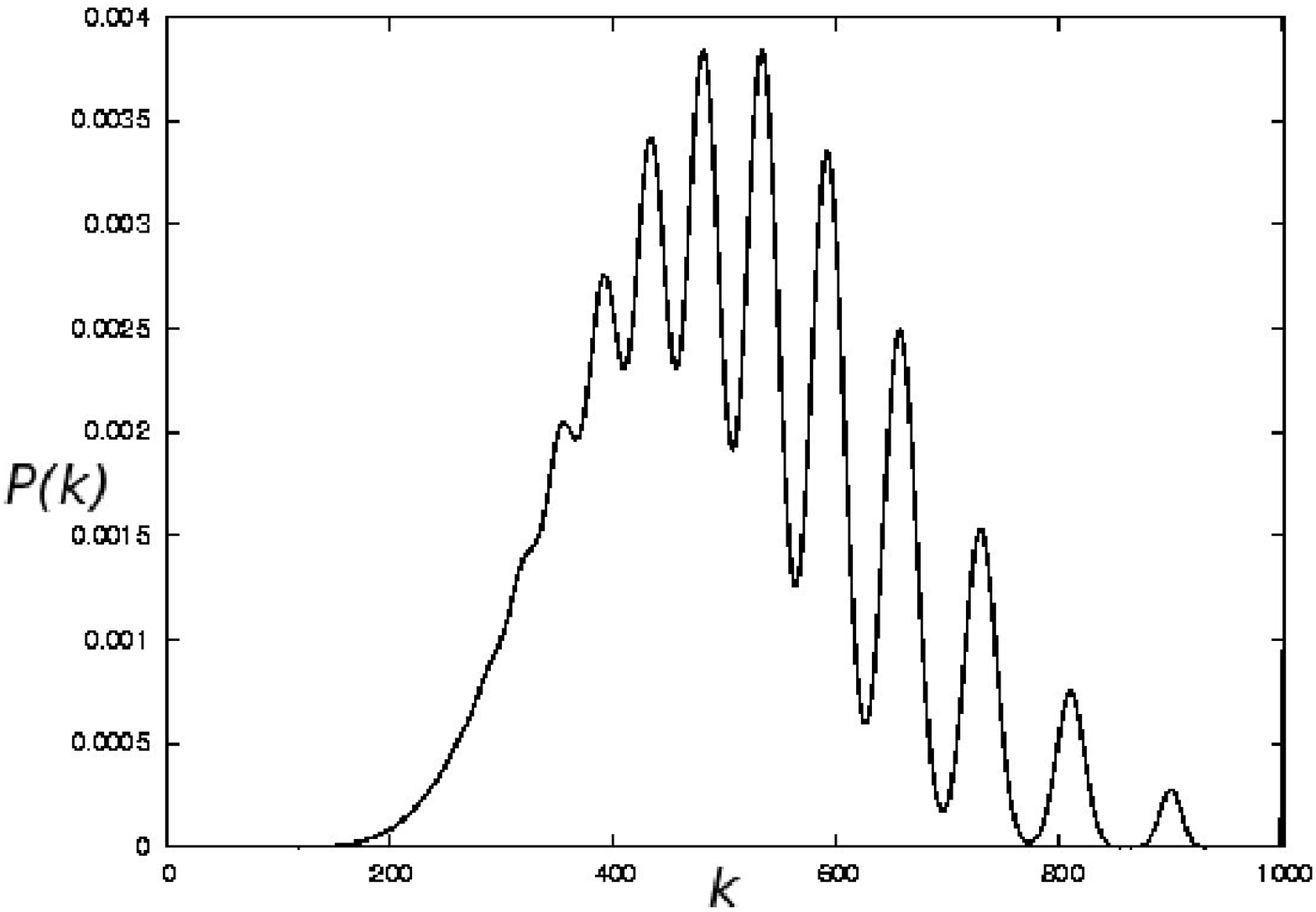}
\caption{Plot of $P(k)$ for $P_{d}=0.5,q=q^{\prime}=0.1,N=1000,\\p=1-\alpha\slash N=0.993$, and$\rho=0.071773$}
\label{right}
\end{center}
\end{minipage}
\hfill
\end{figure}

We show the profiles of the probability distributions $P(k)$
with $p=\alpha\slash N$ and $p=1-\alpha\slash N$ in Fig.\ref{left}
and Fig.\ref{right}. They show reflection symmetric profiles
($k\leftrightarrow N-k$)and 
have very singular oscillating shapes.
Hereafter, we interpret the oscillating behavior 
based on eq.(\ref{alpha}).
The behavior of $P(k)$ can be understood 
by considering each term in eq.(\ref{PK}).
It is expressed as a summation of  
$\alpha^{p,q,q^{\prime}}_{N,k}(l,m)$.
The difference between $P(k)$
for $p=\alpha\slash N$ and $p=1-\alpha\slash N$
is only observed in the first part of eq.(\ref{alpha}), 
$p^{N-k-m+l}(1-p)^{k-l+m}$.
For $p=\alpha\slash N$,
\begin{equation}
p^{N-k-m+l}(1-p)^{k-l+m}=\Big(\frac{\alpha}{N}\Big)^{N-k-m+l}\Big(1-\frac{\alpha}{N}\Big)^{k-l+m}.
\label{first}
\end{equation}
This suggests that $\alpha_{N,k}^{p,q,q^{\prime}}(l,m)$, which 
contributes significantly to $P(k)$,
should satisfy the condition $k+m-l \approx N$.
From the second line of eq.(\ref{alpha}),
$(1-q^{\prime})^{l(k+m-l)}(1-q)^{m(N-k-m+l)}$,
we see that $l$ should be equal to 0 because $k-l+m\approx N$.
Therefore, $\alpha_{N,k}^{p,q,q^{\prime}}$ with
$l=0$ and $m\approx N-k$ significantly contributes
to $P(k)$.
The third part of eq.(\ref{alpha}),
$(1-(1-q)^{N-k-m+l})^{k-l}(1-(1-q^{\prime})^{k+m-l})^{N-k-m}$,
has non zero value with the above condition in the limit $N\to \infty$.

We set $l=0$ and $m=N-k-n \hspace*{0.1cm}
( n\slash N\ll 1)$ in eq.(\ref{alpha}).
The probability distribution function can be expressed as
\begin{equation}
\lim_{N\to\infty}P_{p=\alpha\slash
 N}(k)\approx _{N}C_{k}\times\sum_{n\ll N}\alpha_{N,k}^{p=\alpha\slash N,q,q'}(l=0,m=N-k-n),
\end{equation}
where
\begin{eqnarray}
\nonumber
\alpha_{N,k}^{p=\alpha\slash N,q,q'}(l=0,m=N-k-n)&=&_{N}C_{k}\times _{N-k}C_{N-k-n}\Big(\frac{\alpha}{N}\Big)^{n}\Big(1-\frac{\alpha}{N}\Big)^{N-n}\\
&&\ \ \times(1-q)^{(N-k-n)n}(1-(1-q)^{n})^{k}.
\label{Plm}
\end{eqnarray}

Instead of $k$, we use a normalized variable $x\equiv \frac{k}{N}$
and express $\alpha_{x}^{\alpha,q,q'}(n)
=\alpha_{N,k}^{p=\alpha\slash N,q,q'}(l=0,m=N-Nx-n)$.
The function ${}_{N}C_{N x}\cdot  \alpha_{x}^{\alpha,q,q'}(n)$
has a very narrow profile, and the position of the peak $x_{n}$ is 
given by the condition
$\partial {}_{N}C_{N x}\cdot \alpha_{x}^{\alpha,q,q'}(n)
\slash\partial x=0$ at $x=x_{n}$.
We get
\begin{equation}
x_{n}(1-q)^{n}=(1-x_{n}-\frac{n}{N})(1-(1-q)^{n}).
\label{cond_l}
\end{equation}

\begin{figure}
\begin{center}
\includegraphics[scale=0.4]{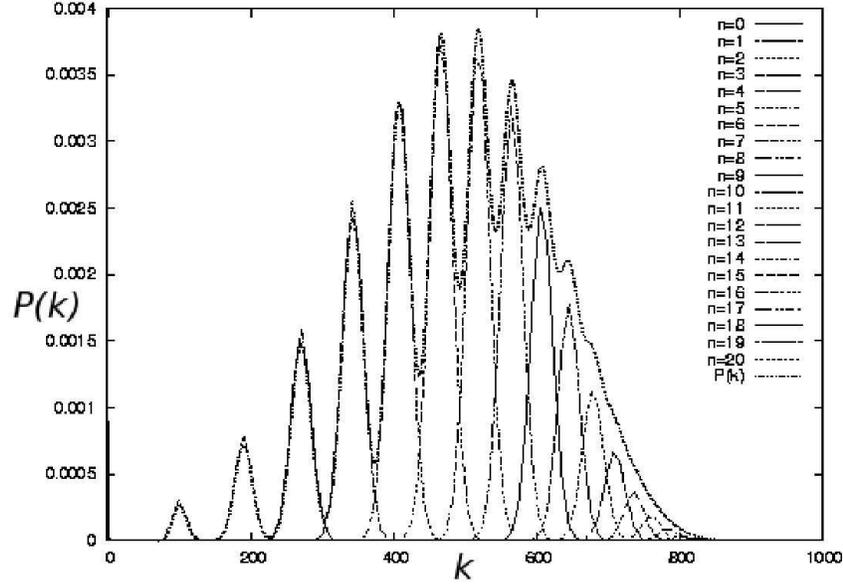} 
\caption{Plot of $P(k)$ of the left solution in Fig.\ref{left} 
and the plot of the normal distribution $p(x\mid n)P_{ini}(n)$ for $0\le n
 \le 20$.}
\label{app_left}
\end{center}
\end{figure}

In the limit $N\to \infty$, 
we obtain the probability density function $p(x)$ as follows.
\begin{equation}
p(x)=\sum_{n=0}\frac{\alpha^{n}e^{-\alpha}}{n!\sqrt{2\pi\sigma_{n}^{2}}}
\exp\Big(-\frac{(x-x_{n})^{2}}{2\sigma_{n}^{2}}\Big),
\label{app_Px}
\end{equation}
where
\begin{eqnarray}
x_{n}&=&1-(1-q)^{n},\\
\sigma_{n}^{2}&=&\frac{1}{N}(1-q)^{n}(1-(1-q)^{n}).
\end{eqnarray}
$p(x)$ can be expressed as
\begin{eqnarray}
p(x)&=&\sum_{n=0}p(x\mid n)P_{ini}(n),   \label{decomp} \\
p(x\mid
 n)&=&\frac{1}{\sqrt{2\pi\sigma_{n}^{2}}}
\exp\Big(-\frac{(x-x(n))^{2}}{2\sigma_{n}^{2}}\Big),\\
P_{ini}(n)&=&\frac{\alpha^{n}e^{-\alpha}}{n!}.
\end{eqnarray}
$P_{ini}(n)$ is the
probability of the occurrence of $n$ internal bad obligors and
it is a Poisson probability function.  
$p(x\mid n)$ is the resulting probability density after $n$ bad
companies appear and infections occur from them.  
 By the decomposition of eq.(\ref{decomp}), it is easy to 
understand the oscillating behavior of $p(x)$. In Fig.\ref{app_left}, 
$p(x)$ is clearly decomposed into the product of the normal
distribution $p(x\mid n)$ and $P_{ini}(n)$.

We comment on De Finetti's theorem, which states that 
the joint probability function
of $N$ exchangeable Bernoulli-type variables can be
expressed as a mixture of the binomial distribution function
$\mbox{Bi}(N,p)$ with some mixing function 
$f(p)$.\cite{df,Ki,CT} In the limit $N\to \infty$, 
$\mbox{Bi}(N,p)$ becomes the delta function $\delta(x-p)$ with 
a suitable normalization.
Our model should be expressed by
such a mixture in the $N\to\infty$ limit.
In eq.(\ref{app_Px}), $p(x\mid n)$ becomes $\delta(x-x_{n})$ 
in the limit $N \to \infty$. The mixing function $f(p)$ is estimated
as
\begin{equation}
f(p)=\sum_{n}P_{ini}(n)\delta(p-x_{n}) .
\end{equation}

We can similarly derive $p(x)$ for the right solution 
$(p=1-\alpha\slash N)$. 
The result is
\begin{eqnarray}
p(x)&=&\sum_{n=0}\frac{\alpha^{n}e^{-\alpha}}{n!\sqrt{2\pi\sigma_{n}^{n}}}
\exp\Big(-\frac{(x-x_{n})^{2}}{2\sigma_{n}^{2}}\Big)
=\sum_{n}p(x\mid n)P_{ini}(n),
\label{appr} \\
x_{n}&=&(1-q^{\prime})^{n},\\
\sigma_{n}^{2}&=&\frac{1}{N}(1-(1-q^{\prime})^{n}).
\end{eqnarray}
 The probability of the occurrence of $n$
good obligors is given by the Poisson probability function $P_{ini}(n)$. 
$p(x\mid n)$ denotes the conditional probability density function for $x$. 

\section{Comparison with implied default distribution}
\label{Comp}

\begin{figure}[tb]
\begin{center}
\includegraphics[scale=0.4]{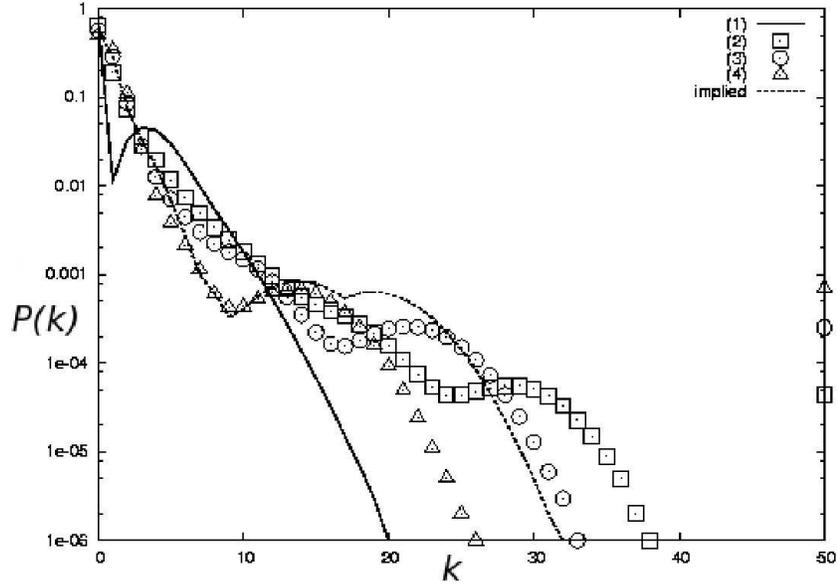} 
\caption{Plot of default probability distribution for the four cases and
 the implied
 distribution on August 30, 2005.}
\label{comp}
\end{center}
\end{figure}

We calibrate  our model based on  the premiums of iTraxx-CJ (Series 2). 
iTraxx-CJ is an equally weighted portfolio 
of $N=50$ credit default swaps (CDSs) on Japanese companies.
The interesting point is that they are divided into several parts 
(called as "tranches"). The tranches have priorities that
are defined by the attachment point $a_{L}$ and detachment point $a_{H}$.
The protection seller agrees to cover all the losses between 
$a_{L}K_{Total}$ and $a_{H}K_{Total}$, where
$K_{Total}$ is the initial total notional of the portfolio. 
That is, if the loss is below $a_{L}K_{Total}$,
the tranche does not cover it. Only when it exceeds $a_{L}K_{Total}$, 
the tranche begins to cover it.  If it exceeds $a_{H}K_{Total}$, 
the notional becomes zero. 
iTraxx-CJ has five tranches and their
attachment  and detachment points are  $\{0\%,3\%\}$,
$\{3\%,6\%\}$,$\{6\%,9\%\}$,
$\{9\%,12\%\}$, and $\{12\%,22\%\}$. In addition, there is an index 
with $\{0\%,100\%\}$. We denote these tranches as
$\{a_{L}^{i},a_{H}^{i}\} (\hspace*{0.1cm}i=1,2,\cdots,6)$ and the 
initial notional as $N_{0}^{i}=(a_{H}^{i}-a_{L}^{i})\times K_{Total}$.

The premium of these tranches depends on the expectation values of the
final notional principal $<N_{T}^{i}>$, where $<\hspace*{0.5cm}>$ denotes
the average over the probability loss functions of the portfolio. 
If there are $k$ defaults, the loss is expressed as the difference of 
the notional principal time and $k$ times the recovery rate $R$, 
set at $35\%$, subtracted from 1.
The final notional principal 
for the $i-$th
tranche is
 \begin{equation}
N_{T}^{i}(k)=
\left\{
\begin{array}{cc}
N_{0}^{i} & k< \lceil \frac{a_{L}^{i}N}{1-R} \rceil \\
a_{H}N-k(1-R)  & \lceil \frac{a_{L}^{i}N}{1-R} \rceil \le k < \lceil
 \frac{a_{H}^{i}N}{1-R} \rceil  \\
0  &  k \ge \lceil \frac{a_{H}^{i}N}{1-R} \rceil . \\
\end{array}
\right.
\end{equation}
Here, $\lceil x \rceil$ denotes the smallest integer greater than $x$.

The premiums of the tranches $s_{i}$ and of the upfront $U_{i}$ are 
determined as follows;
\begin{eqnarray}
&&U_{i}N_{0}^{i}+T <N_{T}^{i}> s_{i}
 e^{-rT}+(N_{0}^{i}-<N_{T}^{i}>)\frac{s_{i}T}{2}e^{-r\frac{T}{2}}
 \nonumber \\
&=&
(N_{0}^{i}-<N_{T}^{i}>)e^{-r\frac{T}{2}} \label{Pre}.
\end{eqnarray}
Here, $r$ is the risk-free rate of interest and we set $r=1\%$.
The left-hand side represents the expected payoff of the contact and 
 the right-hand side represents the expected loss due to defaults.\cite{HW}
Generally, $U_{i}=0$ for $i\ge 2$ and $s_{1}=3\%$.

The premiums include information
about the credit market expectations regarding the probability loss functions.
From these premiums, it is possible to infer the probability 
loss function, which is called the ``implied default distribution.''
It describes the probability of 
$k$ defaults of 50 Japanese  companies. 
We denote it by $P_{imp}(k)$. There are several ways to infer the
implied distribution; in Fig.\ref{comp}, we indicate $P_{imp}(k)$ 
with a dotted line based on the maximum entropy principle.
The default probability $P_{d}$ is estimated as $P_{d}=1.65 \%$,
and the default correlation $\rho$ is $\rho=6.8\%$.
It decreases rapidly for small $k$; for $9 \le k \le 20$, it is
almost constant at $\simeq 0.1\%$. Thereafter, $P_{imp}(k)$ 
rapidly decays to zero. Details of the inference process 
are provided elsewhere.\cite{MKH2} Many probabilistic models
have been proposed to date, 
however they only yield poor fits to the implied distribution.
Here, we calibrate the model parameters $p,q$, and $q'$ and study 
whether or not our model effectively fits $P_{imp}(k)$. 

In the calibration, we 
equate the default probability $P_{d}$ and default correlation $\rho$
of the model with those of the implied ones.
There are three parameters $p,q$, and $q'$ in the model, while there is only one
degree of freedom.
We study $P(k)$ for the following four cases.

\begin{enumerate}
\item Default infection only (left): $q^{\prime}=0.0,q=0.054857$, 
and $p=0.004512$.

\item Recovery infection only (right): $q=0.0,q^{\prime}=0.421050$, 
and $p=0.818175$.

\item Default infection with recovery (right):
   $q=0.001,q^{\prime}=0.563790$, and $p=0.847362$.

\item Default infection with recovery (right):
   $q=0.002,q^{\prime}=0.723940$, and $p=0.864563$.
\end{enumerate}

The $P(k)$ values for the above four cases are shown in Fig.\ref{comp}.
In the first case, the model exhibits only the default infection
mechanism $(q'=0)$, and which is indicated with a solid line. $P(k)$ exhibits 
a sharp valley structure at $k=1$ and then
decreases rapidly to zero at $k \simeq  20$. This profile is clearly different
from that of the implied one. On the other hand, the model with the
right solution $q'>q$ and $p \sim 1$, 
where the recovery effect dominates over the
default infection, the bulk profiles are smooth. They are depicted by the
symbols $\square (q=0.0),\bigcirc (q=0.001)$, and $\triangle (q=0.002)$.
Their profiles are closer to the implied one than that of the
infection only case.
As $q$ increases, the tail becomes short and fat. At $q=0.001$ and 
$q=0.002$, they look similar to the implied one.
We consider the infectious recovery to be important 
to describe the implied default distribution in the framework of
infectious models.

The $P(k)$ values for the right solutions (cases 2, 3, and 4)
have another peak at $k=50$. 
The peak means the probability that all the 50
companies default simultaneously. 
The discrepancy from $P_{imp}(k)$ is not very serious, because
the inference of the default distribution from market quotes depends
on the details of the optimization process. Instead of the entropy
maximum principle, if we use another method, the implied distribution
might have a peak at $k=50$.  

The reason why the peak appears at this position in the infectious models
is that we need to set a large value of $p$ for obtaining the right solution.
The probability that all 50 companies are bad is $p^{50}$, and it
is non zero; in the case, the recovery infection does not occur
because there are no good companies and $P(50)$ remains.
On the other hand, the probability of $k <50$ bad companies and $50-k$ good
companies is $P_{int}(k)={}_{50}C_{k}\cdot p^{k}(1-p)^{50-k}$. 
In this case, $50-k$ good companies support $k$ bad companies, 
and the resulting default number is far less than $k$.
Intuitively, probability $P_{int}(k)$ for $k$ bad companies is shifted 
to the left and it changes to probability $P(k')$ for $k'<k$ defaults.
As $q$ increases, in order to fix $P_{d}$ and $\rho$, we need to
increase $p$ and $q'$. The peak at $k=50$ becomes higher, and 
the shift of $P(k)$ to the left increases. As a result, 
$P(k)$ moves to the left.  

In case 1 where only default infection occurs, the distribution 
shifts to the right in general. In the case of $k=0$ where there are no bad
companies, the probability $(1-p)^{50}$ is maintained. The default infection
does not occur and $P(k)$ has a peak at $k=0$. 

We have also compared the premiums obtained using our model with the real values.
The first row of Table \ref{premiums} lists the quotes for iTraxx-CJ 
(Series 2) on August 30, 2005.
In the second row, we show the premiums derived using our model 
with the parameters of the above four cases. 
We find that case (3) realizes the best match with the real 
premiums.  In the table, we also list the premiums based on the Gaussian
copula loss function, which is a standard model in financial engineering.
\cite{HW} As the model parameter, we use the same $P_{d}$ and $\rho$
values as the implied one's.

\begin{table}
\begin{tabular}{l||cccccc}
\hline
Premiums & $U_{1}$ & $s_{2}$ & $s_3$ & $s_4$ & $s_5$ & $s_6$ (Index) \\ \hline
iTraxx-CJ & 0.131330 & 0.008917 & 0.002850 & 0.002000 & 0.001400 & 0.002208 \\ \hline
IDM 1& 0.056668 & 0.024918 & 0.007940 & 0.001971 & 0.000190 & 0.002203 \\
\hline
IDM 2& 0.107641 & 0.013250 & 0.005137 & 0.002432 & 0.000809 & 0.002202 \\
\hline
IDM 3& 0.133617 & 0.008996 & 0.003695 & 0.002025 & 0.000798 & 0.002203 \\
\hline
IDM 4& 0.096937 & 0.014655 & 0.006025 & 0.002792 & 0.000815 & 0.002203 \\
\hline
Gaussian Model & 0.207827 & 0.004207 & 0.000078 & 0.000000 & 0.000000 & 0.002203 \\
\hline
\end{tabular} 
\label{premiums}
\caption{Premiums for iTraxx-CJ (Series 2) on August 30, 2005, and those derived using our model.}
\end{table}

\section{Concluding remarks and future problems}
\label{Conc}

We have generalized the infectious default model by incorporating an 
infectious recovery effect. 
We have explicitly obtained the default probability 
function $P(k)$ for $k$ defaults as a function of model parameters
$p,q$, and $q'$. 
We have considered the continuous limit and obtained the 
probability density function $p(x)$ for the default ratio 
$x=\frac{k}{N}$. We have understood the oscillating behavior of $p(x)$
by decomposing it, as expressed by eq.(\ref{decomp}). $p(x)$ is expressed as a
superposition of the occurrence of $n$ bad obligors and the following
default infection. The former follows a Poisson distribution and the
latter obeys a normal distribution. The normal distributions have 
narrow peaks of width $\sim \frac{1}{\sqrt{N}}$, which appear in the
oscillating behavior of $p(x)$. We have compared the obtained $P(k)$ with the implied
one $P_{imp}(k)$ inferred from the iTraxx-CJ quotes.

By calibrating the model parameters, the profiles
appear similar with regard to the bulk shape. However, $P(k)$ has a peak at $k=50$. We provide an intuitive
explanation for it.
We note that
the principal features of our model are solvability, 
fitness for the implied distribution, and 
the model can be expressed as a superposition of the Poisson distributions
with only three parameters.
 
In the future, we should study the time evolution of our model.
One possibility is that we prepare an initial configuration of 
$S_{i}(t=0)$, whose time evolution is expressed as
\begin{equation}
S_{i}(t+1)=S_{i}(t)\Pi_{j\neq
 i}(1-Y_{ij}^{\prime}(t)(1-S_{j}(t)))+(1-S_{i}(t))(1-\Pi_{j\neq
 i}(1-Y_{ij}(t)S_{j}(t))).
\end{equation}
$Y_{ij}(t)$ and $Y'_{ij}(t)$ are independent Bernoulli-type variables at each
time $t$, and the configuration of $S_{j}(t)$ is mapped to a new
configuration $S_{i}(t+1)$. In the original problem, the binomial
distribution $\mbox{Bi}(N,p)$
for $X_{i}$ is transformed into a singular oscillating $P(k)$.
We can expect more dynamic and complex behaviors. 
Furthermore, in addition to 
the two-body interaction $Y_{ij},Y'_{ij}$, three-body or 
many body interactions might be interesting; this can be achieved by maintaining the
integrability of the model, to determine the extent, to which such a generalization 
is possible. In the continuous limit, the  model with 
a continuous mixing
function $f(p)$ should be searched. 

The model is defined on the complete graph, where all nodes
are interconnected.
However, in recent times, the industry networks have been extensively studied
and it has been shown that they have complex 
structures.\cite{SW,YF}
The behavior of the model on such realistic networks 
is interesting.
In addition, the relation between this model and
the contact process\cite{MKA,PV,Dc} should be clarified.
Despite the evident similarity of our model to the
contact process, the infectious model proposes  
a new approach to the description of infection.
It may be that we can obtain the attribute of the
contact process from the infectious models by considering some limit.

\section*{Acknowledgment}

One of the authors (A.S.) thanks Prof. K. Kaneko and Dr. K. Hukushima 
for the useful discussions and encouragement.

\end{document}